\documentclass[12pt]{iopart}
\begin{document}

\title{Non-perturbative physics of ${\cal N}=1$ super Yang-Mills theories}

\author{Paolo Merlatti  
\footnote[3]{merlatti@nbi.dk}
}

\address{Nordita Institute, Copenhagen, Denmark}


\begin{abstract}
We discuss some of the recent developments of ${\cal N}=1$ super Yang-Mills theories in the context of the gauge-string correspondence.
\end{abstract}

\section{Introduction}

In recent years ${\cal N}=1$ super Yang-Mills theories have been much investigated and a number of non trivial results about their non-perturbative dynamics have been reached. 

The motivation of this intensive investigation is string theory. Indeed it has emerged that the embedding of gauge theory in string theory (via D-branes) is a powerful tool to explore the fascinating infrared dynamics of the field theory under consideration. Within this framework, various approaches and several line of investigations have been developed, all going under the rather generic name of gauge-string correspondence.
One of the most interesting result that has emerged in this contest is the relation of ${\cal N}=1$ super Yang-Mills to matrix models. This relation has been found following a long path, going through topological string theory, superstring theory and D-branes. 

However, after the matrix model structure of gauge theories has been conjectured in this set-up \cite{dv}, it has also been possible to recover the same results in a purely field theoretical approach \cite{cdsw, csw,dglvz} (for a nice and detailed review see \cite{fer}).

In this talk I will mainly describe the  field theoretical approach  based on the chiral ring structure characterizing ${\cal N}=1$ supersymmetric theories and on a generalization of the Konishi anomaly \cite{cdsw}. Following \cite{cdsw,csw} we will just sketch how things work in the general case, recovering the main results about the solution of the theory. By means of solution here we mean the determination of all the chiral quantities in the supersymmetric vacua. However the solution will be found in a rather implicit form (as it is given in \cite{csw}) and a lot of effort has still to be done to work out physical quantities case by case. We will specialize thus to the case of supersymmetric QCD (SQCD, namely ${\cal N}=1$ super Yang-Mills theories with $N_f$ flavors in the fundamental and anti-fundamental representation of the gauge group), with $SU(N)$ gauge group and we will always be considering the case $N_f<N$. 

For this theory we will determine the structure of the supersymmetric vacua and,  using the general solution we were referring to in the last paragraph, we will compute the low-energy chiral quantity called gaugino condensate. This can sound trivial as this quantity has been already determined using different techniques and it has been argued in \cite{cdsw,mat} that it follows from a more general operator statement holding in any supersymmetric vacuum of the theory. However all the previous computations of this quantity have been done using non-perturbative techniques (such as instantons, monopoles or referring to the non-perturbative information encoded in the parent ${\cal N}=2$ Seiberg Witten curve)\footnote{Not all of these techniques still apply in the presence of matter, the case discussed here}. Also in \cite{cdsw}, the operator statement from which the gaugino condensate follows has been found doing a non-perturbative generalization of a classical result, this generalization coming from a well known instanton computation but being not related to the solution of the theory that has been found in that paper. Thus, it's necessary (or at least reassuring) to see that things work properly for the solution given in \cite{csw}. And moreover it is rather meaningfull that, differently from the other approaches, we will determine the gaugino condensate using only perturbative properties of the theory.

We will turn then to consider again the gauge-string correspondence, that, as I already said, was the original motivation for this field theoretical investigation. We will see thus that, using the knowledge that the gaugino does condensate in a supersymmetric vacuum, the gauge-string correspondence seems to imply some non-perturbative modification of the standard field theoretical description of ${\cal N}=1$ pure Yang-Mills theories. In particular we will be lead to argue that there is some non-perturbative modification of the standard $U(1)_R$ anomaly of these theories. 

\section{The general case}

Let us consider ${\cal N}=1$ supersymmetric $U(N)$ gauge theory with a chiral superfield $\Phi$ in the adjoint representation of the gauge group, $N_f$ matter fields $Q_f$ in the fundamental representation and $N_f$ ($\tilde{Q}_{\tilde{f}}$) in the anti-fundamental one and a generic tree-level superpotential:
\begin{equation}
\label{tree}
W_{tree} = Tr~W(\Phi)+\tilde{Q}_{\tilde{f}}~m_f^{\tilde{f}}(\Phi)Q^f.
\end{equation}
where $W(z)$ is a degree $n+1$ polynomial
\begin{equation}
W(z)=\sum_{k=0}^n \frac{1}{k+1}g_kz^{k+1}\ \ \mbox{and}\ \ W'(z)=g_n\prod_{i=1}^n(z-a_i)
\end{equation}
It's easy to see that in a classical vacuum the gauge group breaks to $\prod_{i=1}^l U(N_i)$ with $l\leq n$.

This theory has a chiral structure that allows us to define the chiral ring. Chiral operators are simply operators that are annihilated by the supersymmetries $\bar{Q}_{\dot{\alpha}}$ of one chirality. The product of two chiral operators is also chiral. Chiral operators are usually considered modulo operators of the form $\{\bar{Q}_{\dot{\alpha}},\ldots\}$. The expectation value of a chiral operator in a supersymmetric vacuum depends only on its equivalence class because the vacuum is annihilated by the supersymmetry generators $\bar{Q}_{\dot{\alpha}}$. The equivalence classes can be multiplied, and form a ring called the chiral ring. A superfield whose lowest component is a chiral operator is called a chiral superfield.

A crucial property of the chiral operators is that the expectation value of a product of chiral operators is independent of each of their positions. Then, using also cluster decomposition, we have
\begin{equation}
\left\langle\prod_IO^I(x_I)\right\rangle~=~\left\langle\prod_IO^I\right\rangle~=~\prod_I\left\langle O^I\right\rangle
\end{equation}

Using some equalities holding in the chiral ring \cite{cdsw}, it is possible to give a complete list of independent single-trace chiral operators. They are $Tr~\Phi^k,\ Tr~\Phi^kW_{\alpha},\ Tr~\Phi^kW_{\alpha}W^{\alpha}$, and $\tilde{Q}_{\tilde{f}}\Phi^kQ^f$ for $f,\ \tilde{f}=1\ldots N_f$. It is possible to write in a compact way all these bosonic operators generating the chiral ring in terms of the gauge invariant quantities:
\begin{eqnarray}
T(z)~=~Tr\frac{1}{z-\Phi}\\ R(z)~=~-\frac{1}{32\pi^2} Tr \frac{W_{\alpha}^{\alpha}}{z-\Phi}\\ M(z)^f_{\tilde{f}}~=~\tilde{Q}_{\tilde{f}}\frac{1}{z-\Phi}Q^f
\end{eqnarray}
Our interest is in the relations that the chiral operators satisfy. These relations are operator statements that hold in any supersymmetric vacuum. Then, if we have enough such relations, we can solve the theory determining all the chiral observables. At a classical level, such relations are simply the classical equations of motion. A quantum generalization of them is given by the perturbative Ward identities that come from the one-loop Konishi anomaly, that is an anomaly in the variation of the superfield $\Phi$, $\delta\Phi=\epsilon\Phi$. But, as in \cite{cdsw}, we can consider the more general variation $\delta \Phi = f(\Phi,{\cal W}_\alpha)$. It turns then out \cite{cdsw} that such variation leads to an anomaly for the current $J_f=Tr~ \bar \Phi e^{ad V} f(\Phi,{\cal W}_\alpha)$, from which the Ward identity follows:
\begin{equation}
\left< \Tr f(\Phi,{\cal W}_\alpha) \frac{\partial
W}{\partial \Phi} \right> =  -\frac{1}{32 \pi^2} \left<
\sum_{i,j}\left[{\cal W}_\alpha , \left[ {\cal W}^\alpha ,
\frac{\partial f}{\partial \Phi_{ij}} \right] \right]_{ji} \right>
\end{equation}
Now, by choosing appropriate variations of the matter field, for instance $\delta \Phi = f(\Phi,{\cal W}_{\alpha }) = R(z)$, we obtain equations for the generating functions $R(z)$ and ${\cal T}(z)$. For example, it is easy to work out the equation for $R(z)$:
\begin{equation}
R^2(z) = W'(z) R(z) + {1\over 4}f(z)\end{equation}
with $f(z)$ a polynomial of degree $n-1$. The solution is 
\begin{equation}
2~ R(z)~=~W'(z)-\sqrt{W'(z)^2+f(z)}
\end{equation}

The appearance of the square-root and the natural requirement that physical quantities (such as $R(z)$) be single-valued imply now that $z$ takes value on the double-sheeted complex plane, a Riemann surface of genus $n-1$. If we define it as $y(z)~=~W'(z)-2~R(z)$, its equation is simply
\begin{equation}
y(z)^2~=~W'(z)^2+f(z).
\end{equation}
The appearance of this Riemann surface is the result of the quantum dynamics of the theory and to find the exact solution corresponds now to solve for the polynomial $f(z)$.

\subsection{The solution}

Let's see now how to find constraints on the generating function ${\cal T}(z)$.
We begin considering that some semiclassical gauge theory data allows to determine its analytic structure  \cite{csw}. In particular it turns out that ${\cal T}(z)$ has only simple poles and the residue at them is integer. Thanks to complex geometry, this is enough to fix it uniquely, apart from a constant.
Indeed, considering the one form $T(z)~dz$, it is possible to find \cite{csw}
\begin{equation}
T(z)~dz ~=~d\ln\psi(z)\end{equation}where\begin{equation}\psi(z)~=~P(z)+\sqrt{P(z)^2-\alpha B(z)}
\end{equation}
with
\begin{equation}
\frac{\alpha}{4}~=~c~\Lambda^{2N-N_f},\ \ c\neq 0,\ \ B(z)~=~Det~ m(z)
\end{equation}
$m$ being the matrix in flavor space appearing in the superpotential and $c$ is an arbitrary dimensionless constant.
Moreover, the polynomial $P(z)$ has to satisfy the following equations:
\begin{eqnarray}
\label{complete}
P(z)^2-\alpha B(z)&=&y(z)^2 H(z)^2\\ W'(z)^2+f(z)&=&y(z)^2
\end{eqnarray}
where the degree of $P(z),\ F(z),\ W(z)$ and $H(z)$ are, respectively, $N,\ 2n,\ n+1$ and $N-n$.

This is the solution of the problem. Indeed by means of these equations now we can fix also $f(z)$, and then all the observable of the theory are determined. Even if this form of the solution is rather implicit and it has to be worked out case by case, this is rather surprising. Indeed we only used the property of the chiral ring and the (generalized) Konishi anomaly. But now all the observable are determined in the full quantum theory, where also non-perturbative effects are taken into account.

It's true that we have still one undetermined constant appearing in front of $\Lambda$, the dynamically generated scale. To fix it simply corresponds to choose a renormalization scheme (it can safely be absorbed in the definition of $\Lambda$).

As a final comment, let me just emphasize that all this perturbative field theoretical analysis has been motivated by string theory results. This is indeed one of the example of the very beautiful and rich interplay that there is between gauge and string theories. In the following, after a simple example, we will go back to string theory and we will see if we can get further insights on the field theory just discussed. 

\section{Example: ${\cal N}=1$ super-QCD}

Let's now use \cite{pm} the generic solution of the last subsection to recover information on ${\cal N}=1$ super Yang-Mills theory (with gauge group $SU(N)$) coupled to $N_f$ massive flavors (always $N_f<N$). We choose a superpotential with a large mass term for the adjoint chiral superfield, so that it is reasonable to integrate it out and to be left with only the fundamental flavors and the glue fields. We then choose the tree-level superpotential to be:
\begin{equation}
W_{\mbox{tree}}~=~\frac{1}{2}M~ Tr\Phi^2~+~\tilde{Q}_{\tilde{f}}~m^{\tilde{f}}_f Q^f
\end{equation}
with $m^{\tilde{f}}_f~=~m_f\delta^{\tilde{f}}_f$ and $M>>m_i$. At energy $E$ such that $m_i<E<<M$, we are then left with ${\cal N}=1$ super-QCD with $N_f$ massive flavors and physical scale \begin{equation}\label{scale}\Lambda_I^{3N-N_f}~=~\Lambda^{2N-N_f}~M^N.\end{equation}

The problem (\ref{complete}) in the case at hand reduces to
\begin{eqnarray}
P_N(z)^2~-~4~\Lambda^{2N-N_f}\prod_{i=1}^{N_f}~m_i~=~F_2(z)~H_{N-1}^2(z)
\label{probl}\\ 
F_2(z)~=~z^2~+~\frac{f_0}{M^2}\end{eqnarray}
where $f_0$ is a constant, related to the gaugino condensate $S$ by \cite{Cachazo:2001jy}:
\begin{equation}
S~=~-\frac{1}{4M}~f
\label{condf}
\end{equation}

This problem can be easily solved \cite{pm} if we consider Chebyshev polynomials (${\cal T}_l(x)$ and ${\cal U}_m(x)$, respectively of first and second kind). Indeed they satisfy the following identity
\begin{equation}
{\cal T}_l(x)^2-1~=~(x^2-1)~{\cal U}_{l-1}(x)^2
\end{equation}

The solution to (\ref{probl}) is thus:
\begin{eqnarray}
P_N(z)~=~2\tilde{\Lambda}^N\eta^N{\cal T}_N\left(\frac{z}{2\eta\tilde{\Lambda}}\right),\hspace{0.5cm}F_2(z)~=~z^2-4\eta^2\tilde{\Lambda}^2,\label{solar}\\ H_{N-1}(z)~=~\eta^{N-1}\tilde{\Lambda}^{N-1}{\cal U}_{N-1}\left(\frac{z}{2\eta\tilde{\Lambda}}\right)
\end{eqnarray}
where 
\begin{equation}
\tilde{\Lambda}^{2N}~=~\Lambda^{2N-N_f}\prod_{i=1}^{N_f}m_i\hspace{2cm}\mbox{and}\hspace{2cm}\eta^{2N}=1;
\end{equation} 
What is most important for us is that the solution (\ref{solar}) also implies: 
\begin{equation}
f~=~-4\eta^2 M^2 (\Lambda^{2N-N_f}\prod_{i=1}^{N_f}m_i)^{1/N},
\label{f}
\end{equation}
Moreover, we know that the equation (\ref{probl}) has been proved to have a unique solution in a very analogous case \cite{Cachazo:2002pr} and that proof is also valid for the case we are considering here. From (\ref{condf}) and (\ref{f}) it is now easy to see that the gaugino condensate is non vanishing and it is equal to
\begin{equation}
S~=~\eta^2 M \left(\Lambda^{2N-N_f}\prod_{i=1}^{N_f}m_i\right)^{1/N}
\label{condtot}
\end{equation}

If we now integrate out the adjoint field $\Phi$, using (\ref{scale}) we easily find
\begin{equation}
S~=~\eta^2 \left(\Lambda_I^{3N-N_f}\prod_{i=1}^{N_f}m_i\right)^{1/N}
\label{condflav}
\end{equation}
This is the expected result in the presence of fundamental flavors \cite{Intriligator:1995au}.

Let's also notice that for the specific case $N_f=0$, (pure ${\cal N}=1$ gluodynamics) we get:
\begin{equation}
S~=~\eta^2\Lambda_I^3
\label{gv}
\end{equation}
and this is in perfect agreement with the expectations of the so called `weak coupling computation' for pure ${\cal N}=1$ gluodynamics (see \cite{Konishi:2003ts} for a related way of getting the same result; see instead \cite{Dorey:2002ik} for a general discussion).

Using standard techniques to integrate in and out various fields, it is possible to relate these results to known superpotentials\footnote{Direct relations between effective superpotentials and matrix models were investigated in \cite{Argurio:2002xv,Demasure:2002sc,Demasure:2003sk}} \cite{Intriligator:1995au}. For example, considering the case $N_f=0$, from (\ref{gv}) and the knowledge of the one-loop $\beta$-function $\beta(g_{YM})=-\frac{3N}{16\pi^2}~g_{YM}^3$, it is easy to recover the Veneziano-Yankielowicz superpotential
\begin{equation}
W_{V.-Y.}~=~S\left(\ln\frac{\Lambda^{3N}}{S^N}+N\right).
\end{equation}
This is the low-energy superpotential for pure ${\cal N}=1$ Yang-Mills theory. It is written in terms of the composite field $S$ and of the dynamically generated scale $\Lambda$. From it we cannot learn nothing about the running of the Yang-Mills coupling constant, whose knowledge we had to assume a priori. 

In the framework of the gauge-string correspondence that motivated all this field theory analysis, it is instead possible to write directly the superpotential in terms of $g_{YM}$ and an ultraviolet cut-off $\Lambda_0$. Then, as we are going to see, it is possible to determine directly the $\beta$-function.

\section{Back to the gauge-geometry correspondence}

The central idea of the gauge-geometry correspondence is \cite{Gopakumar:1998ki,Vafa:2000wi,Cachazo:2001jy} to engineer geometrically the field theory via D-branes wrapped over certain cycles of a non-trivial Calabi-Yau geometry. Thus, the low energy dual arises from a geometric transition of the Calabi-Yau, where the branes have disappeared and have been replaced by suitable fluxes.

The simplest example is pure ${\cal N}=1$ super Yang-Mills theory, engineered on the conifold. Before the transition we have $N$ D5 branes wrapped on the blown-up $S^2$ of a resolved conifold. After the transition we are instead left with $N$ units of Ramon-Ramon flux through the $S^3$ of a deformed conifold.

From the knowledge of the geometry after the transition and the map between the original microscopic field theoretical degrees of freedom and the new geometrical data, it is possible to determine the effective gauge theory superpotential \cite{Cachazo:2001jy}. It turns out to be:
\begin{eqnarray}
\nonumber W_{eff}(S)~=~\frac{8\pi^2}{g_{YM}^2}S+\frac{1}{2}N\Lambda_0^3\sqrt{1-\frac{4S}{\Lambda_0^3}}+\\ \label{fullsupot} +NS\left[\ln\frac{S}{\Lambda_0^3}-2\ln\frac{1}{2}\left(1+\sqrt{1-\frac{4S}{\Lambda_0^3}}\right)\right]\end{eqnarray}
We can now expand it in powers of $\Lambda_0$ (the ultra-violet cut-off at which $g_{YM}$ is evaluated)  and then get:
\begin{equation}\frac{1}{2}N\Lambda_0^3+W_{V.Y.}(S,\Lambda_0,g^2_{YM})+O(\frac{1}{\Lambda_0})
\end{equation}
where
\begin{equation}
W_{V.Y.}(S,\Lambda_0,g^2_{YM})=NS\left(\ln\frac{S}{\Lambda_0^3}+\frac{8\pi^2}{N g_{YM}^2}-1\right)
\label{vy}\end{equation}

From this superpotential and the knowledge of gaugino condensation that we have from the last paragraph, it is possible now to determine the $\beta$-function of the theory.

Indeed, we can minimize (\ref{vy}) with respect to the gaugino bilinear superfield $S$ and impose that the result is the gaugino condensation:
\begin{equation}
S=\Lambda_0^3~ {\mbox e}^{-\frac{8\pi^2}{N g_{YM}^2}}~~=~\star\Lambda^3
\end{equation}
Differentiating this relation we can now get the $\beta$-function
\begin{equation}
\beta (g_{YM})=\frac{\partial g_{YM}}{\partial\ln\frac{\Lambda_0}{\Lambda}}=~-\frac{3N}{16\pi^2}~g_{YM}^3.
\end{equation}
This is the right $\beta$ function in the Wilsonian scheme, where it receives contributions only at one loop.

We can now apply the same procedure also to the full superpotential (\ref{fullsupot}):
\begin{equation}
S=\frac{\Lambda_0^3}{4\cosh^2\frac{4\pi^2}{N g_{YM}^2}}~~=~\star\Lambda^3
\end{equation}
from which we can read the $\beta$-function
\begin{equation}
\beta (g_{YM})=\frac{\partial g_{YM}}{\partial\ln\frac{\Lambda_0}{\Lambda}}=~-\frac{3N}{16\pi^2}g_{YM}^3~\frac{1+ {\mbox e}^{-\frac{8\pi^2}{N g_{YM}^2}}}{1- {\mbox e}^{-\frac{8\pi^2}{N g_{YM}^2}}}
\end{equation}
This result seems to imply that there are non-perturbative correction to the perturbative running of $g_{YM}$. In a supersymmetric theory this means that there are non-perturbative modification also to the $U(1)_R$ anomaly. 

This is not the only case in which the gauge-gravity correspondence suggests that such non-perturbative effects are present. Indeed also in other cases, where the supergravity solution is explicitely known (the Klebanov Strassler solution \cite{ks} and the Maldacena Nu$\tilde{n}$ez one \cite{mn}), it has been emphasized \cite{dlm,bm,i} that the gravitational description seems to include non-perturbative configurations (the so-called fractional instantons) that affect the running of $g_{YM}$.

This analysis would then imply that there are non-perturbative contributions to the anomaly. If this would be the case, the Konishi anomaly itself (the starting point of the field theory analysis we briefly discussed here) should probably be modified. Then also the solution of the gauge theory discussed in this talk should be modified to take into account the non-perturbative corrections. 

This is just one example of how the interplay between gauge and string theories can be very profitable.

\end{document}